\documentclass[namedreferences]{solarphysics}

\usepackage[optionalrh,natbib]{spr-sola-addons}

\usepackage{graphicx}                    
\usepackage{courier}                    
\usepackage{amssymb}                    
\usepackage{bm}
\usepackage{color}                       
\usepackage{url}                         
\usepackage[figuresright]{rotating}

\ifx \doiurl \undefined \def \doiurl#1{\href{http://dx.doi.org/#1}{\url{#1}}}\fi
\ifx \adsurl \undefined \def \adsurl#1{\href{http://adsabs.harvard.edu/abs/#1}{\url{#1}}}\fi

\newcommand{\aap}{    {\it Astron. Astrophys.}}

\newcommand{\apj}{    {\it Astrophys. J.}}

\newcommand{\mnras}{  {\it Mon. Not. Roy. Astron. Soc.}}
\newcommand{\nat}{    {\it Nature}}

\newcommand{\pasj}{   {\it Pub. Astron. Soc. Japan}}

\newcommand{\solphys}{{\it Solar Phys.}}

\newcommand{\vel}{\mbox{{\boldmath$v$}}}  
\newcommand{\br}{\mbox{{\boldmath$r$}}} 
\newcommand{\id}{\mbox{$\rm d$}} 
\newcommand{\bk}{\mbox{{\boldmath$k$}}}	
\newcommand{\sbk}{\mbox{{\boldmath\scriptsize$k$}}} 
\newcommand{\sbr}{\mbox{{\boldmath\scriptsize$r$}}} 

\usepackage{solaheader}	
\newcommand{\solareceiveddate}{1 November 2011} 
\newcommand{\solaaccepteddate}{3 November 2012}

\begin{document}

\begin{article}

\begin{opening}

\title{Testing Helioseismic-Holography Inversions for Supergranular Flows Using Synthetic Data}

%
\author{D.E.~\surname{Dombroski$^1$}\sep
        A.C.~\surname{Birch$^{1,2,3}$}\sep
        D.C.~\surname{Braun$^{1}$}\sep
        S.M.~\surname{Hanasoge$^{2,4}$}      
       }

\runningauthor{D.E.~Dombroski \textit{et al.}}
\runningtitle{Testing Helioseismic Holography Inversions}

                     \institute{$^1$NWRA, CoRA Division, Boulder, CO, USA \\
                     $^2${Max-Planck-Institut f\"{u}r Sonnensystemforschung, 37191 Katlenburg-Lindau, Germany} \\
                     $^3${Georg-August-Universit\"{a}t, Institut f\"{u}r Astrophysik, 37077 G\"{o}ttingen, Germany} \\
                     $^4${Department of Geosciences, Princeton University, Princeton, NJ 08544, USA} \\
                     A.C. Birch: \href{mailto:birch@mps.mpg.de}{birch@mps.mpg.de} \\
               }

\begin{abstract}
Supergranulation is one of the most visible length scales of solar convection and has been studied extensively by local helioseismology.  We use synthetic data computed with the Seismic Propagation through Active Regions and Convection (SPARC) code to test regularized-least squares (RLS) inversions of helioseismic holography measurements for a supergranulation-like flow.  The code simulates the acoustic wavefield by solving the linearized three-dimensional Euler equations in Cartesian geometry.  We model a single supergranulation cell with a simple, axisymmetric, mass-conserving flow.

The use of simulated data provides an opportunity for direct evaluation of the accuracy of measurement and inversion techniques.  The RLS technique applied to helioseismic-holography measurements is generally successful in reproducing the structure of the horizontal flow field of the model supergranule cell.  The errors are significant in horizontal-flow inversions near the top and bottom of the computational domain as well as in vertical-flow inversions throughout the domain.  We show that the errors in the vertical velocity are due largely to cross talk from the horizontal velocity.
\end{abstract}

%
\keywords{Interior, Convection Zone; Supergranulation; Helioseismology, Inverse Modeling; Helioseismology, Direct Modeling; Waves, Propagation; Velocity Fields, Interior}

\end{opening}

%
\section{Introduction}
	\label{sec.intro} 

Wave travel times are influenced by the properties of the medium in which they propagate; helioseismic holography \citep{Lindsey1990,Lindsey2000} can be used to measure travel-time shifts, and is therefore useful for probing the interior structure of the Sun \citep{Gizon2005,Gizon2010}.  

Supergranulation is a visible pattern on the solar surface often interpreted as a scale of solar convection.  Despite extensive research by local helioseismology, there remains substantial uncertainty regarding the characteristic scales and fundamental processes responsible for the creation and sustainment of supergranulation patterns \citep[\textit{e.g.}][]{Gizon2005,Retord2010b}.  

Supergranulation cells were first detected by Hart (\citeyear{Hart1954,Hart1956}); observations and analysis of data over a period of five decades have revealed average cell sizes in the range 15\, --\, 30 Mm \citep[\textit{e.g.}][]{Hart1956,Leighton1962,Duvall1980,Hathaway1992,Hagenaar1997,DeRosa2004,Hirzberger2008}, with discrepancies in estimated scale due largely to differences in analysis techniques.  Similarly, direct inferences of the velocity field have produced a range of horizontal-flow estimates \citep[\textit{e.g.}][]{Hart1954,Simon1964,Hathaway2002}.  Rieutord \textit{et al.} (\citeyear{Retord2010a}), motivated by observed scale-dependency in surface kinetic-energy distributions, performed a spectral analysis of velocities estimated using granule tracking and Doppler measurements.  They measured a velocity of 300 m$\,$s$^{-1}$ at a scale of 36 Mm.  Estimating vertical velocity has proven more difficult due to the noise contribution and low signal strength, particularly near the boundaries of supergranulation cells \citep{Retord2010b}.  Results from SOHO/MDI data \citep{Hathaway2002} and \textit{Hinode}/SOT data \citep{Retord2010a} produced estimates of 30 m$\,$s$^{-1}$.  \cite{Duvall2010} estimated cell-center vertical flow to be about 10 m$\,$s$^{-1}$ using direct Doppler measurements averaged over 1100 supergranules. 

Several theories have been advanced to explain the development of supergranulation flow patterns at the observed preferential scale, although confirmation or rebuttal of these theories is largely lacking.  An early theory proposed by \cite{Simon1964} attributed supergranular scales to convective instability from the recombination of ionized helium.  More recently, \cite{Retord2000,Retord2001} proposed a model whereby exploding granules trigger large-scale instability of the granular flow, leading to supergranulation.  \cite{Ploner2000} used a 1D model to similarly show that the interaction and merging of individual granular plumes may determine supergranular scales.  \cite{Rast2003} demonstrated that large spatial and long temporal supergranular scales arise in a simplified advective model that simulates the interaction of many small-scale and short-lived granular downflow plumes.  \cite{Crouch2007} were able to simulate correlations between cell size and magnetic activity using a model based on a random-walk approximation to the dispersal and interaction of small-scale magnetic elements at the solar surface.  Qualitatively, they found that this process produced supergranule-like spatial patterns.  

Debate continues regarding the structure of supergranule flows below the solar surface \citep[\textit{e.g.}][]{Duvall1998,Zhao2003,Woodard2007,Sekii2007}, largely due to the difficulties associated with estimating relevant quantities using helioseismology \citep[\textit{e.g.}][]{Gizon2005,Retord2010b}.  These challenges are compounded by the sound speed and density stratification, which effectively decrease resolution and signal-to-noise ratio with depth, making the detection of a return flow elusive.  \citet{Kosovichev1997} first used the time--distance helioseismology technique \citep{Duvall1993} to make measurements of supergranular flow patterns from Doppler measurements.  \citet{Zhao2003} also used time--distance helioseismology to infer supergranule flow fields; they found that horizontal flows could be derived reliably within a few megameters of the surface, but that vertical flows remained uncertain.  They determined supergranulation to be a relatively deep phenomenon with a full pattern depth of up to 15 Mm and a return flow measured below a depth of 5\, ---\, 6 Mm.  \citet{Braun2004}, using helioseismic holography, attributed the measured return flow to signal leakage, and concluded that supergranulation is a relatively shallow phenomenon.  \citet{Woodard2007} used a forward model to perform subsurface inversions of Doppler data, however detection was limited to 4\, --\, 5 Mm below the photosphere.\citet{Jackiewicz2008} used a novel 2+1D optimally localized averaging (OLA) method to study supergranulation in the top few Mm.  \citet{Braun2007} and \citet{Zhao2007} used synthetic data from numerical simulations of wave propagation through supergranule-like flows to test the performance of helioseismic techniques.  The former compared model travel times with those computed from synthetic observations using helioseismic holography.  The latter computed inversions using time--distance helioseismology and compared the results to the modeled flows.  Both studies documented limitations in the ability of the techniques to detect the full extent of the flow fields.  \citet{Svanda2011} tested the OLA inversion technique of \citet{Jackiewicz2012} on synthetic travel times and showed improved ability to infer vertical flow fields. 

In this study, we simulate wave propagation through a simple kinematic model of a supergranule flow pattern and make helioseismic holography measurements from observations of the resulting velocity field.  The aim is to test the quality of regularized least squares \citep[RLS: \textit{e.g.}][in the context of local helioseismology]{Zhao2001,Kosovichev1996} inversions and gain insight into the limitations of this commonly used technique in local helioseismology by comparing the calculations with a known simulated flow field.  We find that the RLS technique applied to helioseismic holography measurements is able to infer some general features of the horizontal field, but fails to infer the vertical field throughout the domain.  Errors in vertical-flow inversions have previously been attributed to cross-talk effects \citep[\textit{e.g.}][]{Zhao2007, Jackiewicz2012}; herein we show the individual contributions from the divergent flow that comprise these effects.

\section{Simulations}
	\label{S-sim}

\subsection{Numerical Algorithm}
	\label{S-comps}
	
We used the SPARC code \citep{Hanasoge2008,Hanasoge2006b,Hanasoge2006a} to solve the three-dimensional linearized Euler equations in Cartesian geometry.  The code computes derivatives in the vertical direction using sixth-order compact finite differences \citep{Lele1992}, while derivatives in the horizontal directions are computed spectrally with periodic boundaries.   An absorbing sponge is used to damp wave reflection at the top and bottom boundaries \citep{Hanasoge2006b}.
		
 Wave excitation is applied 200~km below the photosphere via a source term in the momentum equation \citep{Hanasoge2007}.  The forcing is statistically uniform in the horizontal directions and Gaussian in the vertical direction with FWHM 200~km.  In the spectral domain, the forcing distribution is localized to within the wavenumber range $k=0-2$ rad~Mm$^{-1}$ and has a frequency maximum at 3 mHz.
 
 Two different 24-hour (solar time) simulation cases were performed.  The first case used the supergranule model with velocity vector $\bm{v} = v_x\bm{\hat{x}} + v_y\bm{\hat{y}} + v_z\bm{\hat{z}}$, where $x$ and $y$ define the horizontal coordinate system and $z$ defines vertical distance from the photosphere.  The second simulation case used no background flows, which allowed for application of the ``noise subtraction" technique in the analysis \citep{Werne2004}.

\subsection{Supergranulation Model}
	\label{supergmod}

The static background model is a convectively stabilized \citep[CSM\_A:][]{Schunker2011} variant of Model S \citep{Christ1996}, which specifies the time-invariant components of the density [$\rho$], pressure [$p$], first adiabatic index [$\Gamma_1$], gravity [$g$], and sound speed [$c_s$].  The computational box is a $300^3$ grid, spanning nonuniformly from $z_{\rm bot}= -25$ Mm to $z_{\rm top}=2.5$ Mm in the vertical direction.  The grid spans 100 Mm uniformly in the $x$- and $y$-directions.  

\begin{figure}
\centerline{\includegraphics[width=0.5\textwidth,clip=]{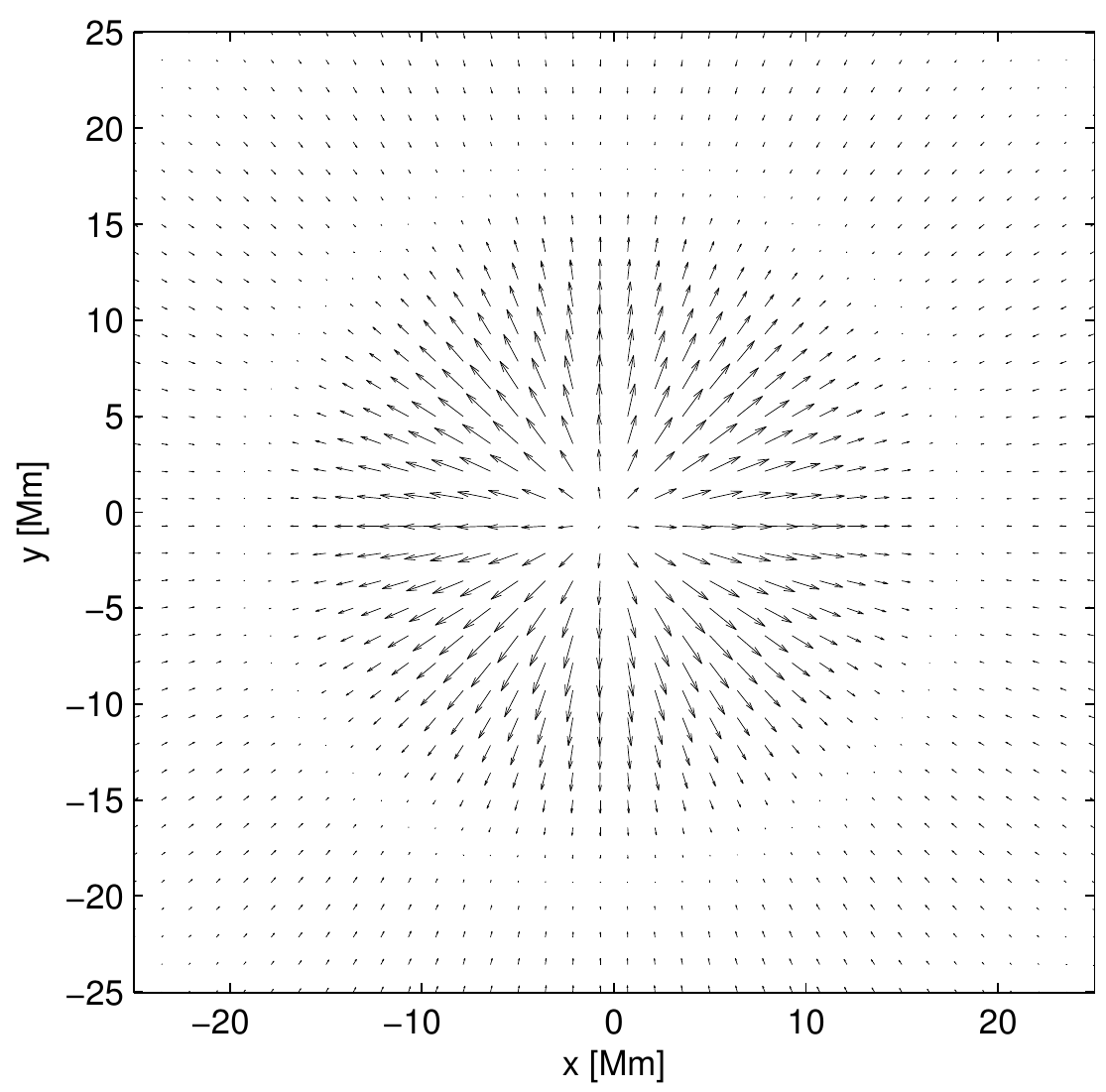}}
\centerline{\includegraphics[width=0.5\textwidth,clip=]{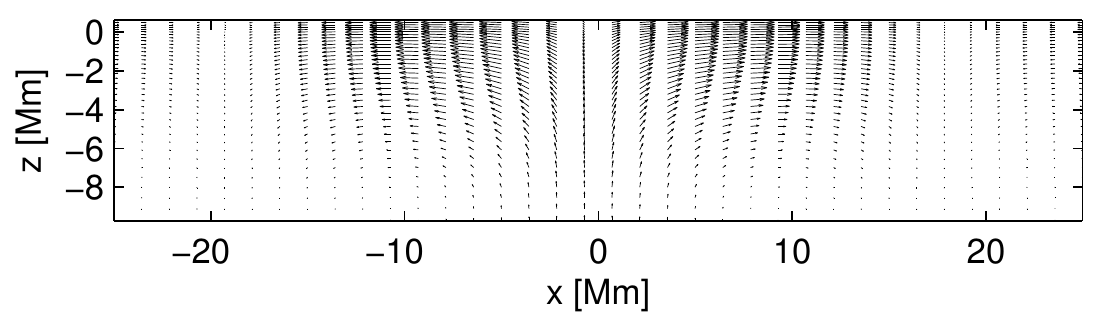}}
\caption{Two-dimensional vector plots of the model supergranulation flow field.  The top panel shows a horizontal slice through the horizontal flow field at the photospheric level ($z=0$) and the bottom panel shows a vertical slice through the flow field at the center of the supergranule ($y=0$).  The full computational domain extends from -50 Mm to +50 Mm in the horizontal directions, where $(x,y) = (0,0)$ corresponds to the center of the supergranulation cell.  Along the photospheric plane, the maximum horizontal and vertical velocities are 250 m s$^{-1}$ and 20 m s$^{-1}$.}
\label{fig.model}
\end{figure}

We model a single supergranule with a simple axisymmetric mass-conserving flow (Figure~\ref{fig.model}) prescribed by the curl of a potential function
   \begin{equation} \label{Eq-superg-def}
     \rho\mathbf{v} = \nabla\times\mathbf{A},
   \end{equation}
   where, for the sake of simplicity, we assume that the potential is separable as
  \begin{equation} \label{Eq-potential-def}
    \mathbf{A} = A_\phi\bm{\hat{\phi}} = \rho(z)h(z) F(r)\bm{\hat{\phi}}
  \end{equation}
with functions $h(z)$ and $F(r)$ defining the vertical and radial dependance of the potential.  The variables $\phi$, $r$, and $z$ denote the angular, radial, and vertical cylindrical coordinates, respectively.  The formulation of Equation~\ref{Eq-superg-def} is chosen so that mass conservation $\nabla\cdot\rho{\bf v}=0$ is automatically enforced.  We chose the radial distribution $F(r)$ to be of the form 
  \begin{equation} \label{Eq-rad-distr}
    F(r) = J_1(kr)\mbox{e}^{-r/R},
  \end{equation}
where $J_1$ is the first order Bessel function of the first kind, $k={2\pi}/{30}$ rad Mm$^{-1}$, and $R=15$ Mm.  This formulation forces decaying reversal of flow direction with increasing radial distance from the axial center.  This functional form is similar to the flow associated with the ``average supergranule'' as seen by, \textit{e.g.} \citet{Duvall2012}.

From Equation~(\ref{Eq-superg-def}), the radial and vertical mass-flux distributions, respectively, are
  \begin{eqnarray}
    \rho v_r &=& - \frac{\partial A_{\phi}}{\partial z},  \label{Eq-rhovr-def}  \\
    \rho v_z &=& \frac{1}{r}\frac{\partial}{\partial r}[rA_{\phi}].  \label{Eq-rhovz-def}
  \end{eqnarray}
  
We chose to model the $v_r$ component of flow as a sum of outflow [$v_{\mbox{out}}$] and inflow [$v_{\mbox{in}}$] Gaussian distributions, such that  
  \begin{equation}
    v_r = v_{\mbox{out}} + v_{\mbox{in}} = \bigl[\alpha_1\exp(-\frac{(z-z_1)^2}{D_1^2}) - \alpha_2\exp(-\frac{(z-z_2)^2}{D_2^2})\bigr]F(r),
  \end{equation} 
where $\alpha_1=250 \mbox{ m s}^{-1},\alpha_2=6.27 \mbox{ m s}^{-1},z_1=0.2 \mbox{ Mm}, z_2=-15\mbox{ Mm},\mbox{and } D_1=D_2=5\mbox{ Mm}$.  Using mass conservation, the coefficient $\alpha_2$ is calculated from the known density distribution $\rho$ and $v_{\mbox{out}}$ with chosen coefficient $\alpha_1$.  The potential function $A_{\phi}$ can then be calculated directly from 
   \begin{equation} \label{Eq-A-int}
    A_{\phi}(r,z) = -\int^{z}_{z_{\rm bot}} \rho(\zeta) v_r(\zeta) \mathrm{d}\zeta,
  \end{equation}
In obtaining Equation~(\ref{Eq-A-int}), we set the integration constant equal to zero since we want the vertical velocity $v_z$ to be zero at the lower boundary (see Equation~\ref{Eq-rhovz} below).

With $A_{\phi}$ fully prescribed, the vertical flow [$v_z$] is calculated from Equation~(\ref{Eq-rhovz-def}): 
  \begin{equation} \label{Eq-rhovz}
    \rho v_z = \frac{A_{\phi}}{r} + \frac{\partial A_{\phi}}{\partial r} = \rho(z)h(z)\bigl[\frac{F(r)}{r} + \frac{\partial F(r)}{\partial r} \bigr].
  \end{equation}

\noindent{}Note that due to the large density gradient with solar depth, outflow velocities near the photosphere are much larger than return flow velocities at greater depth in order to conserve mass flux $\rho \mathbf{v}$ (Figure~\ref{fig.model}).

\section{Travel-time Measurements}
\label{sec.traveltimes}

Starting from the vertical velocity taken at a height of 200 km, we used helioseismic holography \citep{Lindsey2000} to measure center--quadrant local-control correlations.  These correlations are analagous to center--quadrant time--distance correlations that have traditionally been used to measure time--distance travel times \citep[\textit{e.g.}][]{Gizon2005,Gizon2010}.  We used the quadrant geometry and ridge filters described by \citet{Braun2008}. In addition to the ridge filters, we also applied band-pass frequency filters to isolate particular ranges in frequency.  The filters had central frequencies of 2.75, 3.00, 3.25, 3.50, 3.75, 4.00, 4.25, and 4.50~mHz and all had widths of 0.25 mHz. We use $\delta\tau_x(\br)$ ($\delta\tau_y(\br)$) to denote $x$ ($y$) direction center--quadrant travel-time shifts and $\delta\tau_{\rm{oi}}=\delta\tau_{\rm{out}}(\br)-\delta\tau_{\rm{in}}(\br)$ to denote out minus in center--quadrant travel-time shifts, where $\delta\tau_{\rm{out}}(\br)$ ($\delta\tau_{\rm{in}}(\br)$) represent travel-time shifts of the observed outgoing (incoming) waves.

\begin{figure} 
\centerline{\includegraphics[width=0.5\textwidth,clip=]{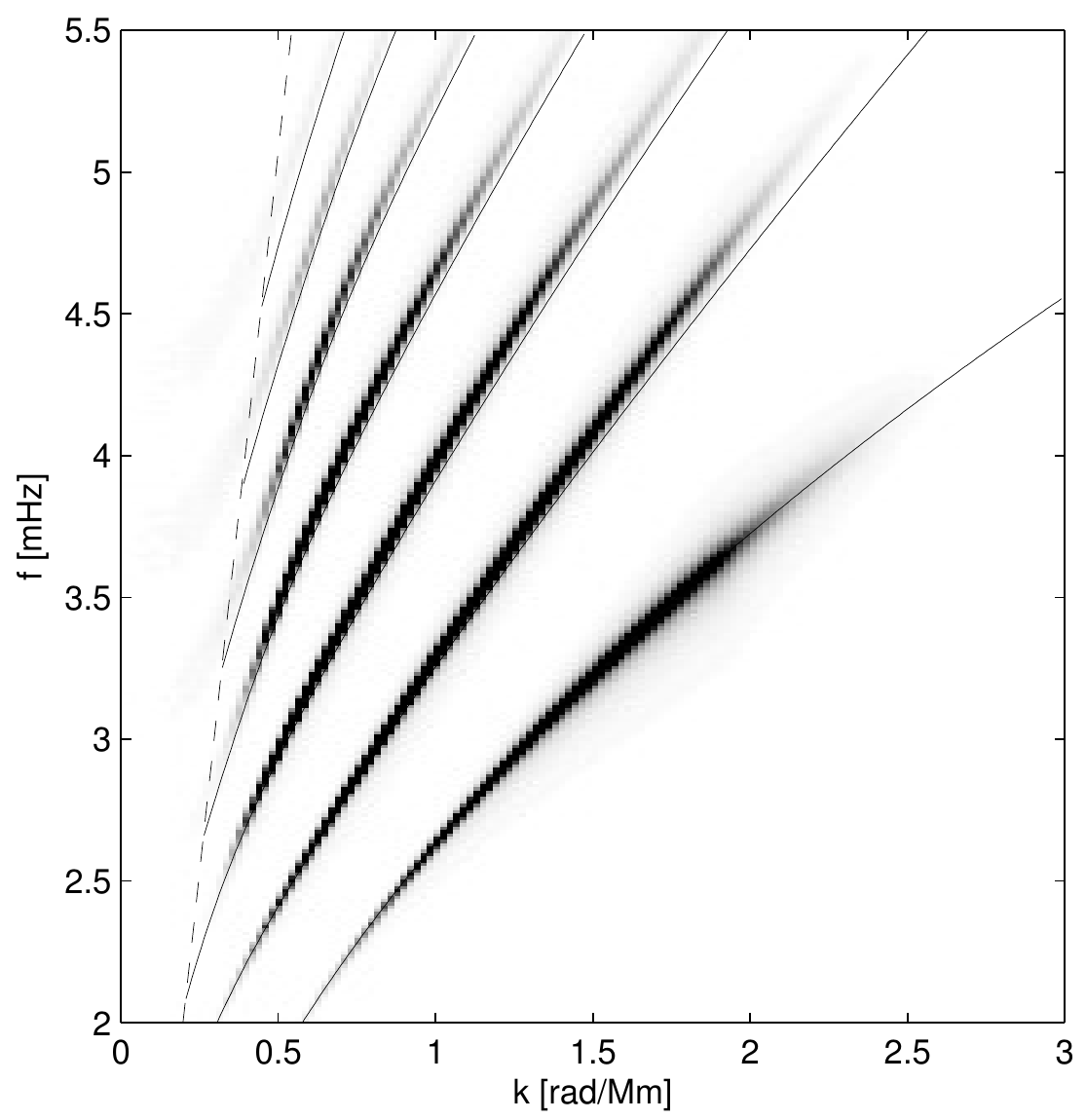}}
\caption{Power spectrum for waves propagated through the CSM\_A background model \citep{Schunker2011}.  The solid black lines denote the computed quiet Sun Model S mode frequencies and the dashed line denotes the horizontal phase speed corresponding to the sound speed at the bottom of the computational box.}
\label{fig.power}
\end{figure}

\begin{figure}
\centerline{\includegraphics[width=4in,clip=]{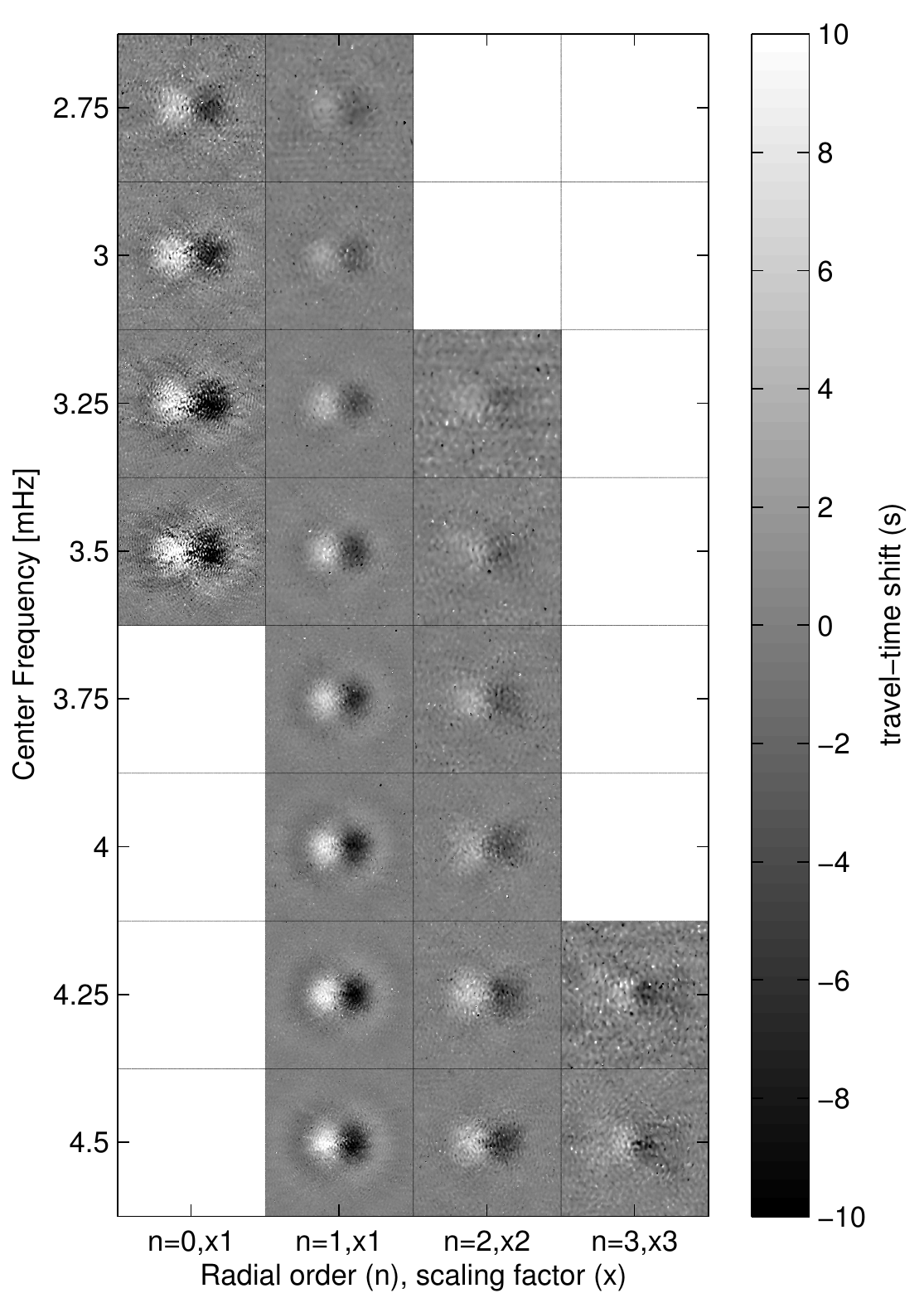}}
\caption{Noise subtracted center--quadrant travel-time differences $\delta\tau_x(\br)$ calculated using ridge filters and 0.25 mHz wide band-pass frequency filters for the supergranulation simulation.  The label at the bottom of each column denotes the radial order of the filter (\textit{e.g.}\ $n=0$ is the $f$-mode, $n=1$ is the $p_1$-mode, \textit{etc.}) and the factor by which the mode data has been scaled (\textit{e.g.} $\times$3 means the data has been multiplied by a factor of three).}
\label{fig.dt_x_ridge_sg}
\end{figure}

\begin{figure}
\centerline{\includegraphics[width=4in,clip=]{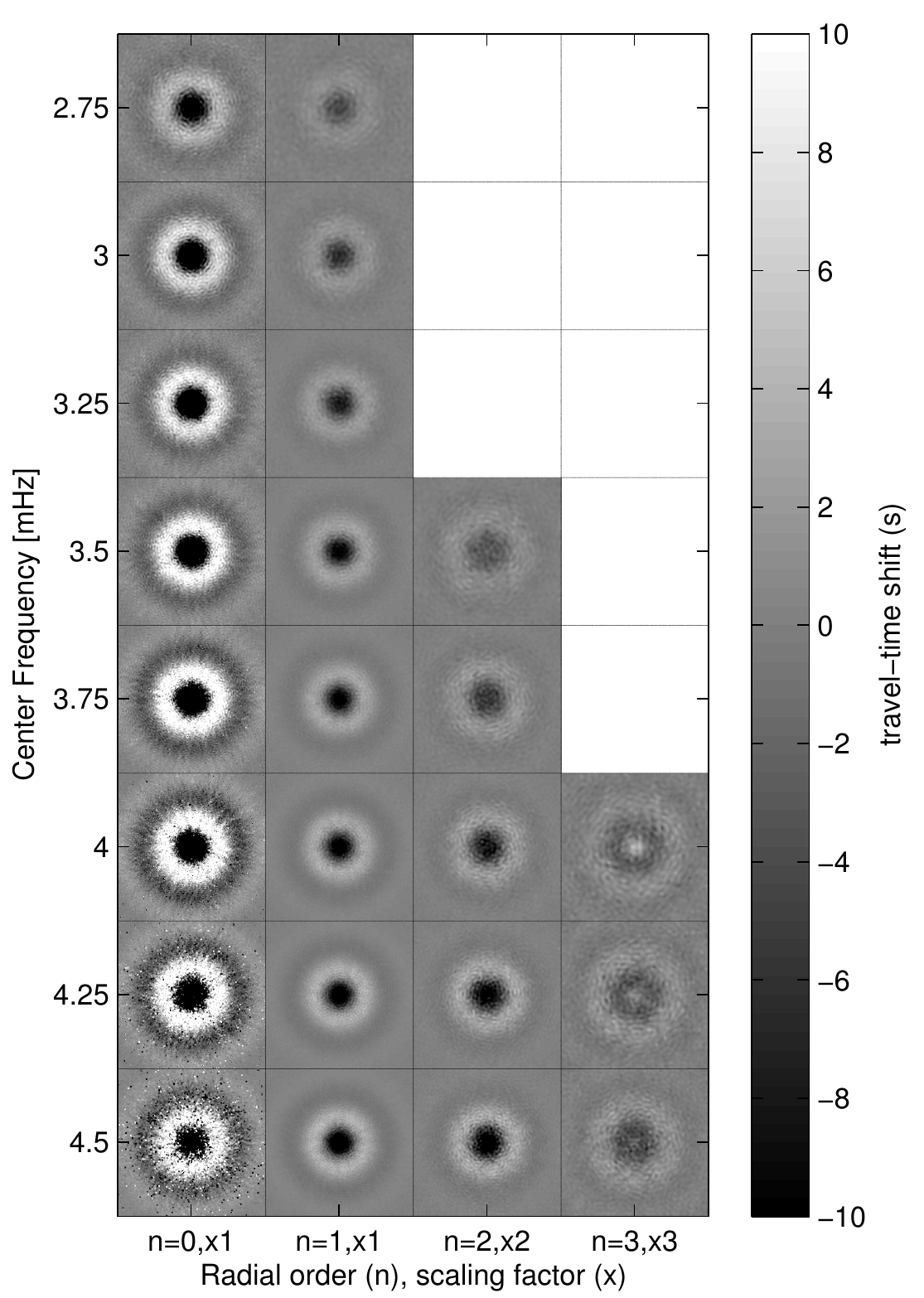}}
\caption{Noise subtracted center--annulus travel-time differences $\delta\tau_{\rm{oi}}$ calculated using ridge filters and 0.25 mHz wide band--pass frequency filters for the supergranulation simulation.  The label at the bottom of each column denotes the radial order of the filter (\textit{e.g.}\ $n=0$ is the $f$-mode, $n=1$ is the $p_1$-mode, \textit{etc.}) and the factor by which the mode data has been scaled (\textit{e.g.} $\times$3 means the data has been multiplied by a factor of three).  The signal-to-noise ratio of the travel-time differences are generally greater than those of $\delta\tau_x$ (Figure~\ref{fig.dt_x_ridge_sg}).}
\label{fig.dt_oi_ridge_sg}
\end{figure}

Figure~\ref{fig.power} shows the power spectrum of the vertical velocity at a height of 200 km above the photosphere in the simulation without any imposed flows.  The resonant frequencies of the simulation are close to those from model~S.  As a result, we apply the ridge filters and holography Green's functions of \citet{Braun2008} without modification.  The travel-time measurements shown in Figures~\ref{fig.dt_x_ridge_sg} and~\ref{fig.dt_oi_ridge_sg} were produced by subtracting the travel-time measurements from the simulation without flows \citep[i.e., ``noise subtraction"; ][]{Werne2004}.

Figure~\ref{fig.dt_x_ridge_sg} shows center--quadrant travel-time differences [$\delta\tau_x$] for a range of radial orders and frequencies for the case where the synthetic data is produced from the supergranulation model.  The flow produces travel-time shifts that generally decrease in amplitude with increasing radial order and increase in amplitude with increasing frequency (i.e., the travel-time shifts decrease with increasing horizontal phase speed).  The spatial pattern is much the same in all cases and is what would be expected given the flow geometry in Figure~\ref{fig.model}.

Figure~\ref{fig.dt_oi_ridge_sg} shows the $\delta\tau_{\rm{oi}}$ travel-time differences resulting from the supergranulation model.  The general trend of the travel-time shifts with radial order and frequency band is consistent with that in Figure~\ref{fig.dt_x_ridge_sg}, although the $\delta\tau_{\rm{oi}}$ measurements feature a larger signal-to-noise ratio.

\section{Flow Inversions}
\label{Inversion_text}

The inverse problem is to estimate the flow field defined by the supergranule model given the travel-time measurements in Section~\ref{sec.traveltimes}.  Here we show some example inversions in which we use the $\delta\tau_x$, $\delta\tau_y$, and $\delta\tau_{\rm{oi}}$ maps to infer the subsurface flows $\vel$.  

We carry out the inversions of the travel-time maps with a noise level corresponding to what we would expect for an average over 100 supergranules measured for 24 hours each.  To generate the input travel-time maps we start from the noise-subtraction travel-time maps (Figures.~\ref{fig.dt_x_ridge_sg} and~\ref{fig.dt_oi_ridge_sg}) and add noise (computed from the simulations carried out in the flow-free reference model) with an amplitude reduced by $1/\sqrt{100}$.

We use kernels computed in the Born approximation using the method of \citet{Birch2007} with erratum \citet{Birch2011}, and account for the holography Green's functions using the approach of \citet{Birch2010}.  The vector-valued kernel functions ${\bf K}$ relate the flows in the interior to the travel-time maps,
\begin{equation}
\delta\tau_i(\br) =\iiint {\bf K}_i(\br'-\br,z) \cdot \vel(\br',z) \; \id\br'\id z \; .  
\label{eq.kernels}
\end{equation}  
In the above equation, we have separate maps and kernel functions for each combination of ridge filter and frequency range.

\begin{sidewaysfigure}
\includegraphics[width=7in,clip=]{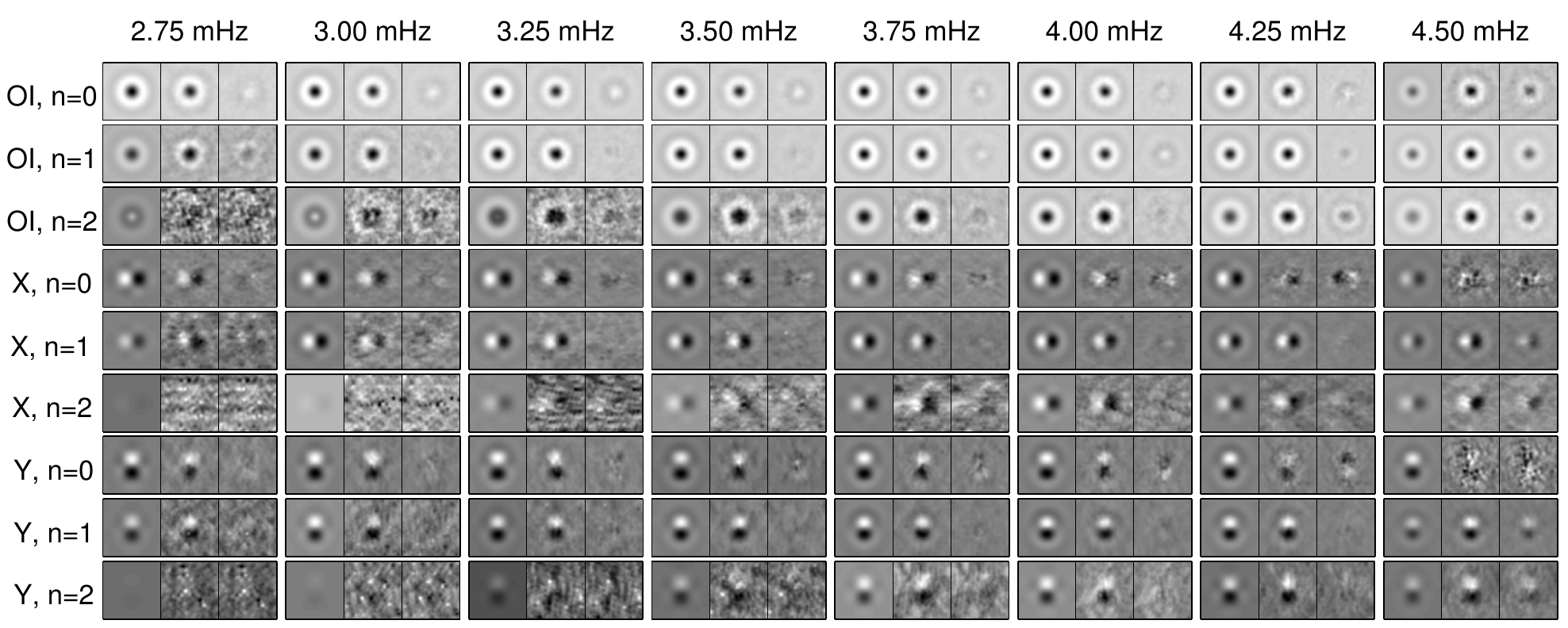}
\caption{Travel-time measurements compared with the expectations from a forward model based on the true flow in the simulation (Equation.~[\ref{eq.kernels}]).  Each panel corresponds to a particular combination of frequency filter, radial order, and measurement geometry.  The three images in each panel contain the forward model associated with the true flow (left), the measurement from the simulation (middle), and the residual (right), respectively. As discussed in the text, the forward model becomes less accurate at high frequencies.  The maps have been smoothed with a gaussian filter ($\sigma=$ ten pixels) to more clearly elucidate the differences, and the panes are scaled independently.}
\label{fig.extra}
\end{sidewaysfigure}

Figure~\ref{fig.extra} compares the measured travel-time shifts with the travel-time shifts that we would expect from Equation~(\ref{eq.kernels}) evaluated for the known flow in the simulation.  At low frequencies there is general, although not perfect, agreement.  At high frequencies, especially 4.5~mHz, there are very significant differences.  This disagreement may be a result of the simplifications made in the calculation of the sensitivity kernels, for example the neglect of the sponge layers in the simulation.  We discuss this issue in more detail in Sections~\S\ref{Results_text} and~\ref{sec.discussion}.

In order to discretize Equation~(\ref{eq.kernels}), we parameterized the flow [$\vel$] as
\begin{equation}
\vel(\br,z) = \sum_{j=1}^N\sum_{\sbk} {\bf a}_j(\sbk)\phi_j(z) e^{i\sbk\cdot\sbr}
\end{equation}
where the sum over $j$ is over a total of $N$ basis functions $\phi_j(z)$ and the sum over $\bk$ is a two-dimensional inverse FFT on the grid that the travel-time maps are measured on (the simulation domain is periodic; as a result, the travel-time maps are periodic as well).  Here we choose the basis functions $\phi_j(z)$ to be one when $z_j <  z \leq z_{j+1}$ and zero otherwise.  The points ${z_i}$ consist of a grid of 56 points that are equally spaced in acoustic depth and cover the range from 0.5 Mm above the photosphere to 10 Mm below the photosphere.  The grid spacing in $k$-space is $h_k = 0.063 \mbox{ rad Mm}^{-1}$.

\begin{figure}
\centerline{\includegraphics[width=4in,clip=]{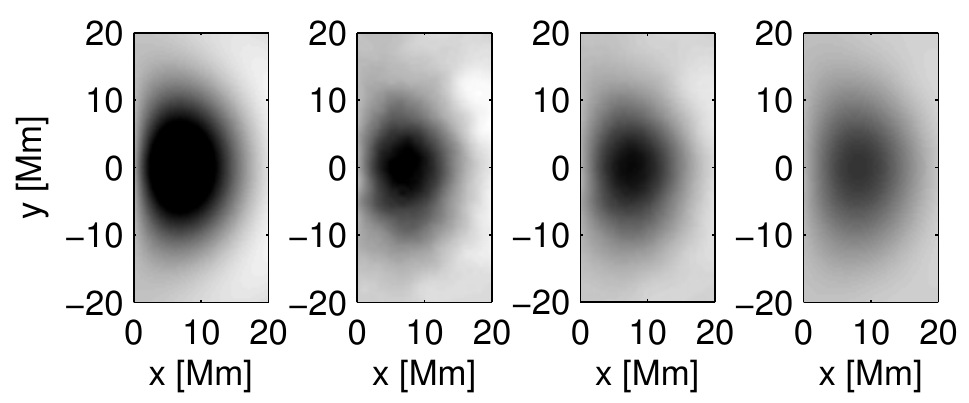}}
\caption{Demonstration of different regularization applied to the $v_x$ inversion with corresponding decrease in resolved structure.  The left panel shows the true $v_x$ at $z=-0.8$~Mm. From left to right, the remaining panels show the inversion results for $v_x$ for regularization parameters ($\lambda^2_1,\lambda^2_2/h_x^2$) = ($1.0\times 10^{-8}$, $5.66\times 10^{-5}$)~Mm$^{-1}$~(m$\,$s$^{-1}$)$^{-2}$ (second panel from left), ($\lambda^2_1,\lambda^2_2/h_x^2$) = ($3.24\times 10^{-8}$ ,$1.76 \times 10^{-4}$)~Mm$^{-1}$~(m $\,$s$^{-1}$)$^{-2}$ (second panel from right),  ($\lambda^2_1,\lambda^2_2/h_x^2$) = ($1.76\times 10^{-7}$, $9.31\times 10^{-4}$)~Mm$^{-1}$~(m \, s$^{-1}$)$^{-2}$ (right panel), where $h_x=1/3$~Mm is the grid spacing.  The upper (lower) bound on the color scale in each panel is 150 (-50) $\mbox{m s}^{-1}$, with black (white) corresponding to positive (negative) values.}
\label{fig.compare_regularization}
%

\end{figure}

We used the MCD approach \citep{Jacobsen1999} to split the full inversion problem into small one-dimensional ($z$-only) inversion problems at each horizontal wavenumber. These problems we solved using RLS \citep[\textit{e.g.}][in the context of local helioseismology]{Zhao2001,Kosovichev1996} with a regularization term given by the depth integral (at each horizontal wavevector $\bk$) of the quantity $v_x^2+v_y^2+10v_z^2$; the factor ten was chosen to reflect the preconception that vertical velocities are on average smaller than horizontal flows.  Following \citet{Couvidat2005}, we used a $\bk$-dependent regularization parameter with $\lambda^2=\lambda_1^2 + \lambda_2^2\|\bk\|^2$, where $(\lambda_1,\lambda_2)$ are fixed parameters.  As the regularization parameters are increased, the amplitude of the large-scale flow in the inversion result is reduced.  We chose the parameters by inspection for each inversion based on a compromise between the desire to average out small-scale noise and the goal of retaining the large-scale flow features.  Figure~\ref{fig.compare_regularization} shows a comparison of different regularization parameters applied to an inversion result for a portion of the $v_x$ velocity field.

The inversion results presented herein reflect an estimate of the mean flow averaged over 100 supergranules for a 24-hour window. Application to real data necessitates consideration of the effective averaging window relative to the time scale over which the coherent structure evolves; a typical supergranule lifetime is estimated to be a day or more \citep{Worden1976,Hirzberger2008}.

\section{Results}
\label{Results_text}

\begin{figure}[ht]
\begin{minipage}{0.48\textwidth}
\centering
\includegraphics[width=\textwidth,clip=]{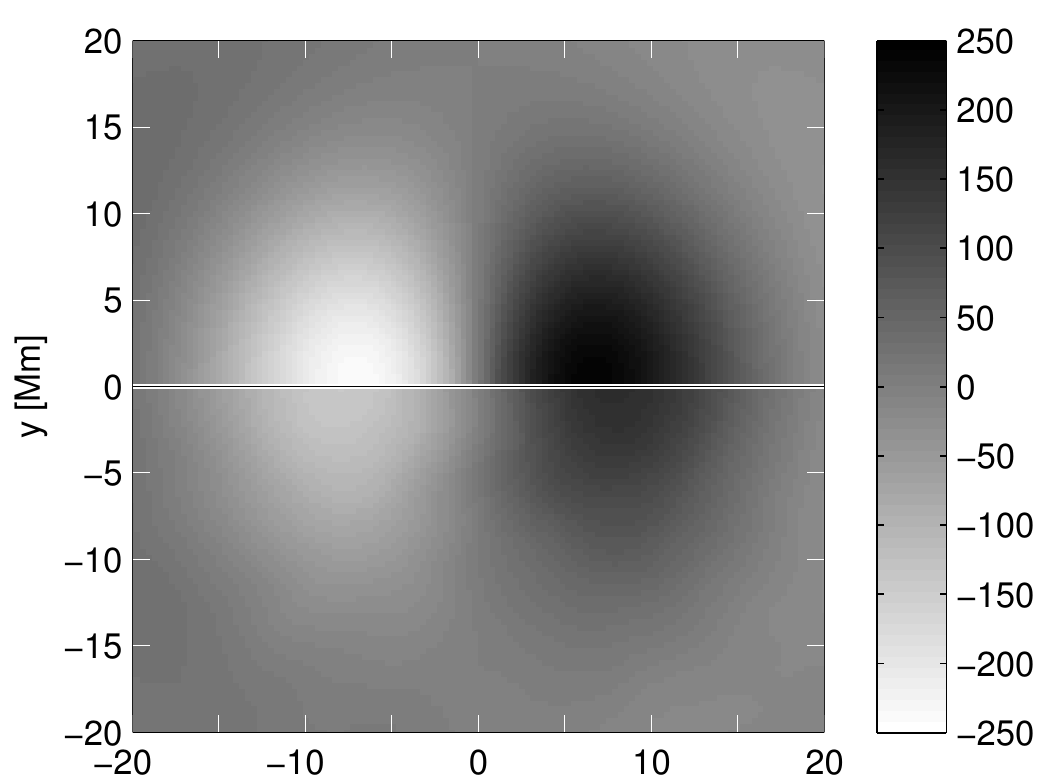}
\includegraphics[width=\textwidth,clip=]{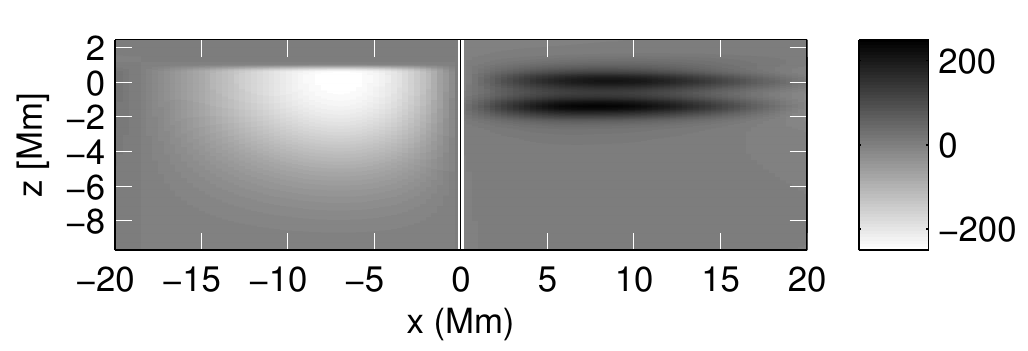}
\end{minipage}
\hspace{0 cm}
\begin{minipage}{0.5\textwidth}
\centering
\includegraphics[width=\textwidth,clip=]{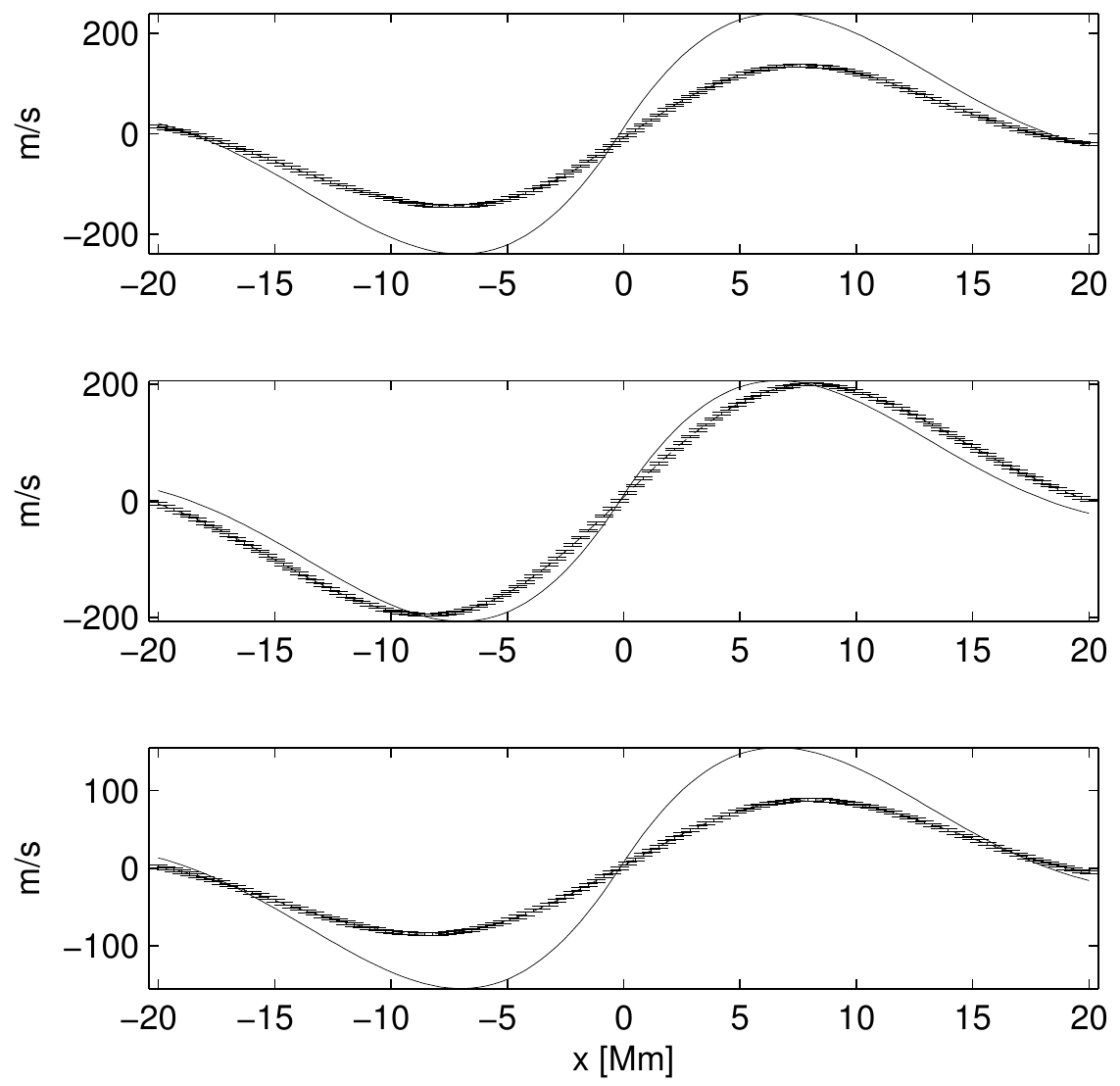}
\end{minipage}
\caption{Result of helioseismic inversion for the $x$-component of the flow in the supergranule simulation using all available $n=0$, $n=1$, and $n=2$ measurements.  The top-left plot compares the true (top panel) and inverted (bottom panel) $v_x(x,y)$ flow at $z=-0.8$ Mm.  The bottom-left plot compares the true (left panel) and inverted (right panel) $v_x(x,z)$ flow at $y = 0$ Mm.  The plots on the right compare the centerline true (solid line) and inverted (dashed line with error bars) $v_x$-velocity at $z=-0.8$ Mm (top panel), $z=-2.0$ Mm (middle panel), and $z=-3.2$ Mm (bottom panel).  The horizontal axes are cropped relative to the full computational domain.  The inversion produces velocity magnitudes that are comparable to the true signal, however misses the velocity signal at the top and lower portion of the domain.}
\label{Vx_all}
\end{figure}

\begin{figure}[ht]
\begin{minipage}{0.48\textwidth}
\centering
\includegraphics[width=\textwidth,clip=]{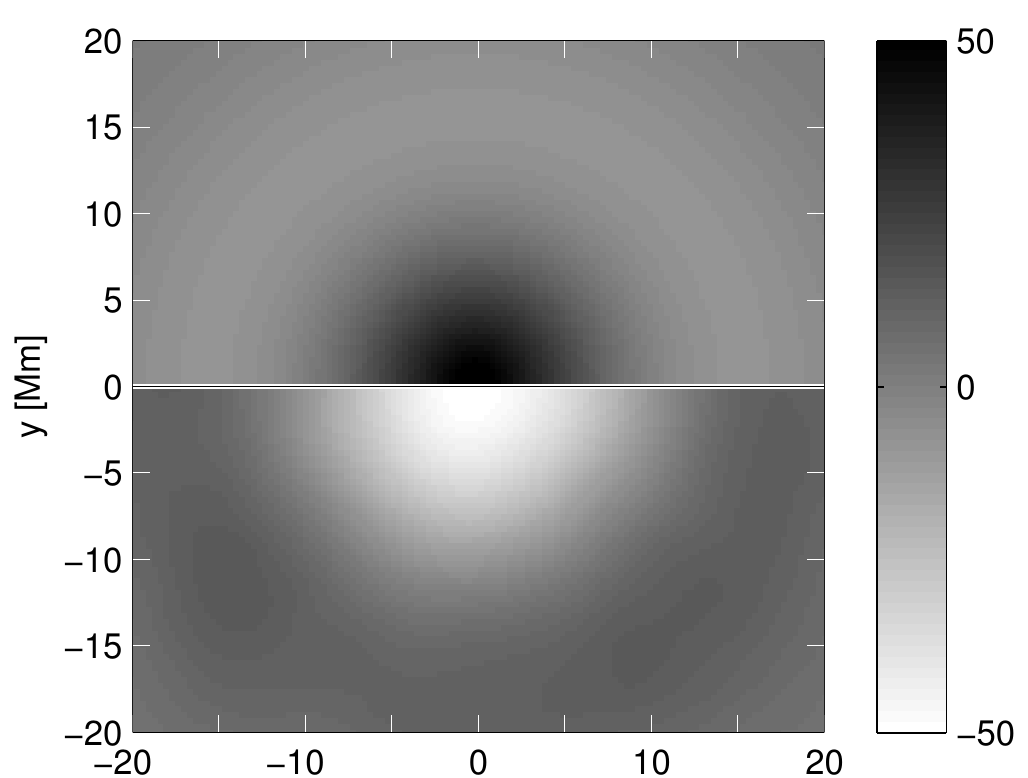}
\includegraphics[width=\textwidth,clip=]{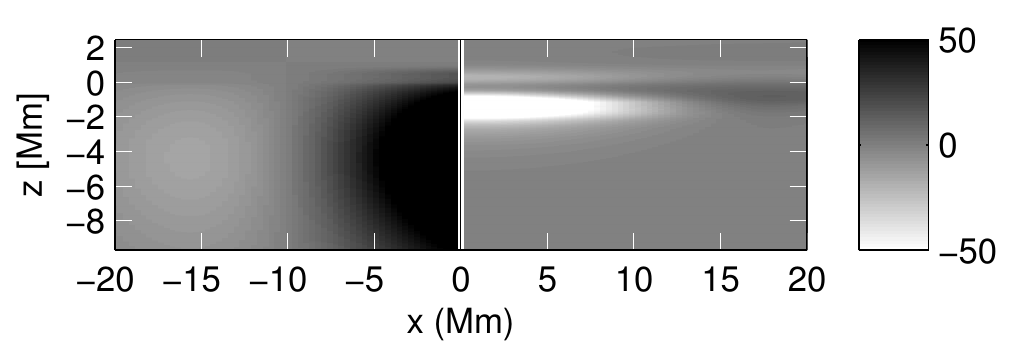}
\end{minipage}
\hspace{0 cm}
\begin{minipage}{0.5\textwidth}
\centering
\includegraphics[width=\textwidth,clip=]{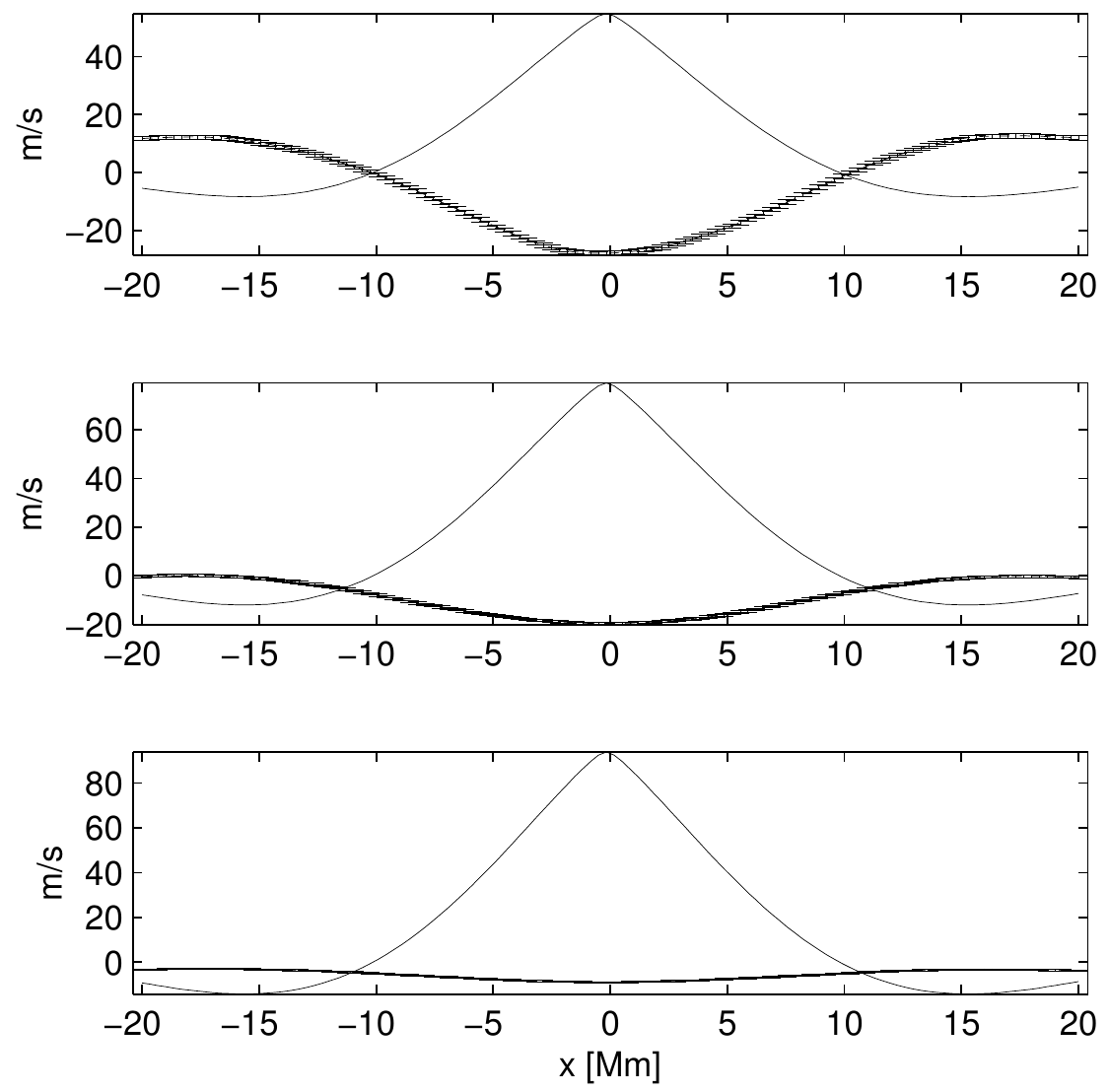}
\end{minipage}
\caption{Result of helioseismic inversion for the $z$-component of the flow in the supergranule simulation using all available $n=0$, $n=1$, and $n=2$ measurements.  The top left plot compares the true (top panel) and inverted (bottom panel) $v_z(x,y)$ flow at $z=-0.8$ Mm.  The bottom left plot compares the true (left panel) and inverted (right panel) $v_z(x,z)$ flow at $y = 0$ Mm.  The plots on the right compare the centerline true (solid line) and inverted (dashed line with error bars) $v_z$-velocity at $z=-0.8$ Mm (top panel), $z=-2.0$ Mm (middle panel), and $z=-3.2$ Mm (bottom panel).  The horizontal axes are cropped relative to the full computational domain.  The inversion fails to qualitatively capture the true $v_z$ flow throughout the physical domain.  As we will discuss later, this is due to cross-talk effects between signals.}
\label{Vz_all}
\end{figure}

Figures~\ref{Vx_all} and \ref{Vz_all} show inversion results using all available $n=0$, $n=1$, and $n=2$ measurements for ($\lambda_1^2,\lambda^2_2/h_x^2$) = ($3.24\times 10^{-8}$, $1.69\times 10^{-4}$)~Mm$^{-1}$~(m$\,$s$^{-1}$)$^{-2}$ compared to the true $v_x$ and $v_z$ velocity components in the supergranule simulation, respectively.  The structure of the $v_x$-inverted fields in Figure~\ref{Vx_all} shares qualitative agreement with the true flow, although misses the signal at the top and bottom (below about $z=-4$ Mm) of the physical domain.  The underestimation of flow magnitude in the lower layers is consistent with prior work \citep[\textit{e.g.}][]{Zhao2001}.  The $v_z$ inversion in Figure~\ref{Vz_all} fails to qualitatively capture the true flow throughout the physical domain.  As we will discuss later, this is due to cross-talk effects between components of the velocity.

We tested the importance of including the $n=2$ measurements by recomputing the inversions shown in Figures~\ref{Vx_all} and~\ref{Vz_all} using only the $n=0$ and $n=1$ measurements.  There was no discernible visual difference in the inversion results for the horizontal and vertical velocity due to exclusion of the $n=2$ measurements, which indicates that the RLS technique is relatively insensitive to the addition of higher radial order measurements. 
 
We tested the effect of noise in the measurements by recalculating the flow inversions using only the $n=0$ and $n=1$ measurements shown in Figures~\ref{fig.dt_x_ridge_sg} and \ref{fig.dt_oi_ridge_sg}, which excludes the noisiest measurements (\textit{i.e.} the whited-out boxes in the grid in Figures~\ref{fig.dt_x_ridge_sg} and \ref{fig.dt_oi_ridge_sg}) that were included in the inversion calculations shown in Figures~\ref{Vx_all} \& \ref{Vz_all}.  There was no discernible visual difference in the inversion results for the horizontal and vertical velocity due to exclusion of the noisiest measurements, which supports the conclusion that the RLS technique is relatively insensitive to changes in the number and quality of measurements used in the inversion calculations.   

\begin{sidewaysfigure}
\centerline{\includegraphics[width=8in,clip=]{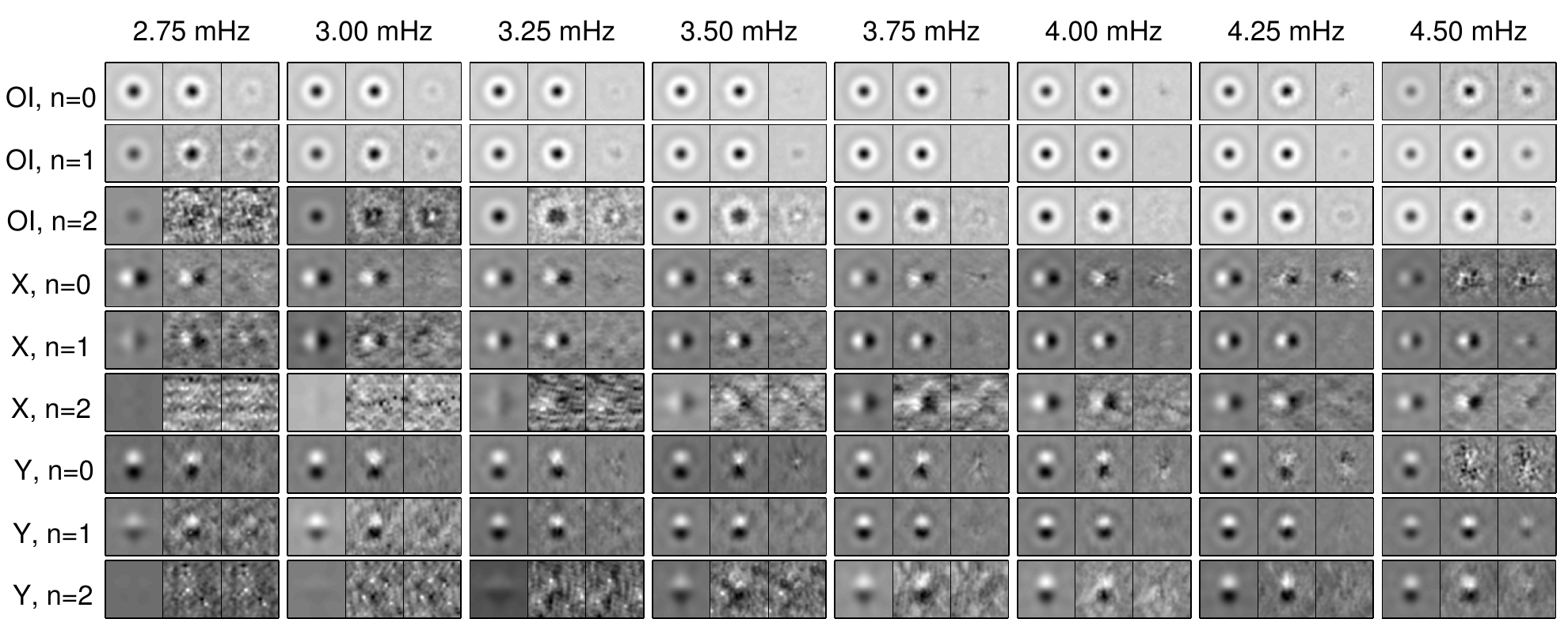}}
\caption{Travel-time measurements calculated from the forward model compared to measurements from the simulation.  Each panel corresponds to a particular combination of frequency filter, radial order, and measurement geometry.  The three images in each panel contain the forward model associated with the inversion result (left), the measurement from the simulation (middle), and the residual (right), respectively.  All of the $n=0$ and $n=1$ measurements generally reflect the structure of the model supergranule flow; the $n=2$ measurements reflect a significant noise contribution.  For all radial orders, the $\delta \tau_x$ and $\delta \tau_y$ measurements contain more noise than the  $\delta \tau_{\rm{oi}}$ measurements.  The maps have been smoothed with a gaussian filter ($\sigma =$ 10 pixels) to more clearly elucidate the differences, and the panes are scaled independently.}
\label{fig.dt_diff_maps}
\end{sidewaysfigure}

We calculated center-quadrant travel-time differences from Equation~(\ref{eq.kernels}) using the kernels and the inferred flow from Figures~\ref{Vx_all} and~\ref{Vz_all} and compared these to the measurements from the simulation.  Figure~\ref{fig.dt_diff_maps} shows these comparisons and the calculated residual (difference between the two) for every combination of frequency filter, radial order, and measurement geometry.  There is generally good correspondence between the perturbations in the $n=0$ and $n=1$ measurements; however, some of the $\delta \tau_x$ and $\delta \tau_y$ measurements, as well as all of the $n=2$ measurements, contain significantly more noise.  In some of the higher-frequency measurements, there is noticeable structure in the residual. This may be due to inaccuracies in the kernels (see Figure~\ref{fig.extra}).  We tested the sensitivity of the inversion to the high-frequency measurements by repeating the inversion using the same regularization parameters, but with the measurements at 4.25~mHz and 4.5~mHz removed.  The main impact of removing these measurements was a reduction in the amplitude of the inferred vertical flow by about a factor of four.

\begin{figure}
\centerline{\includegraphics[width=4.75in,clip=]{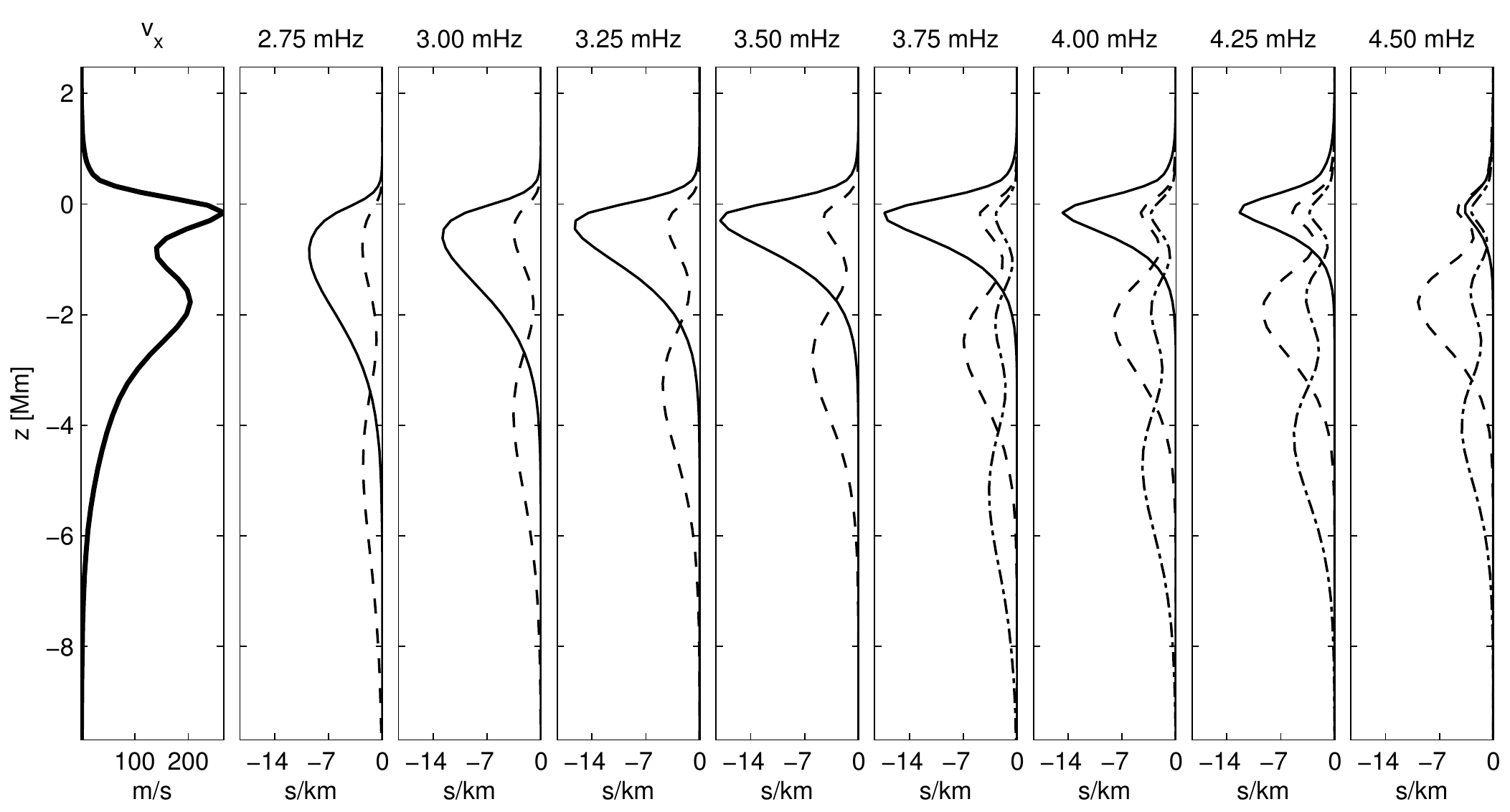}}
\caption{Horizontally integrated $\delta\tau_x$ kernels grouped by radial order and frequency, normalized by their respective rms noise level.  Shown are $n=0$ kernels (solid line), $n=1$ kernels (dashed lines), and $n=2$ kernels (dot--dashed lines).  The left-most panel shows a slice through the $v_x$ inversion result for reference.  The figure demonstrates that the lobed nature of the $v_x$ inversion is due to the form of the kernels, which exhibit localized minima and maxima as a function of $z$.  Furthermore, the sensitivity of the $n=2$ kernels is generally less than the $n=0$ and $n=1$ kernels, which means that use of higher radial-order measurements in the inversion calculations will have decreasing effectiveness on the result.}
\label{fig.kernel_profiles}
\end{figure}

Our findings are generally consistent with those of \citet{Zhao2007}, who used the time--distance method and kernels computed from the ray approximation to perform inversions of simulated supergranule-scale convection.  They were able to obtain reasonable inversion results only down to about 4 Mm below the surface, although they credit this to a limit in the largest annulus radius used.  Here, the sensitivity functions for many of the travel-time measurements extend well below $z=-4$ Mm.  This suggests that, with a sufficiently small noise level, it will be possible to detect flows below this depth (although with a spatial resolution that decreases with increasing depth).  The structure of the horizontal velocity inversion results can be better explained by observing the form of the sensitivity kernels.  Figure~\ref{fig.kernel_profiles} contains a vertical slice through the $v_x$ inversion (left-most panel) and horizontally integrated $\delta\tau_x$ kernels grouped by radial order and frequency, and normalized by the rms noise level.  The $n=0$ kernels have a single band of sensitivity with a maximum located in between 0 and -1 Mm.  The $n=1$ kernels have two regions of sensitivity which explains why the $v_x$ inversion has two local maxima in the vertical direction.  The $n=2$ kernels have several regions of sensitivity; however, inclusion of the $n=2$ measurements did not significantly effect the performance of the inversions because the signal-to-noise ratio of the $n=2$ measurements is about a factor of two lower than that of the $n=0$ and $n=1$ measurements (Figure~\ref{fig.dt_oi_ridge_sg}).

\begin{figure}
 \centerline{\hspace*{0.015\textwidth}
               \includegraphics[width=0.515\textwidth,clip=]{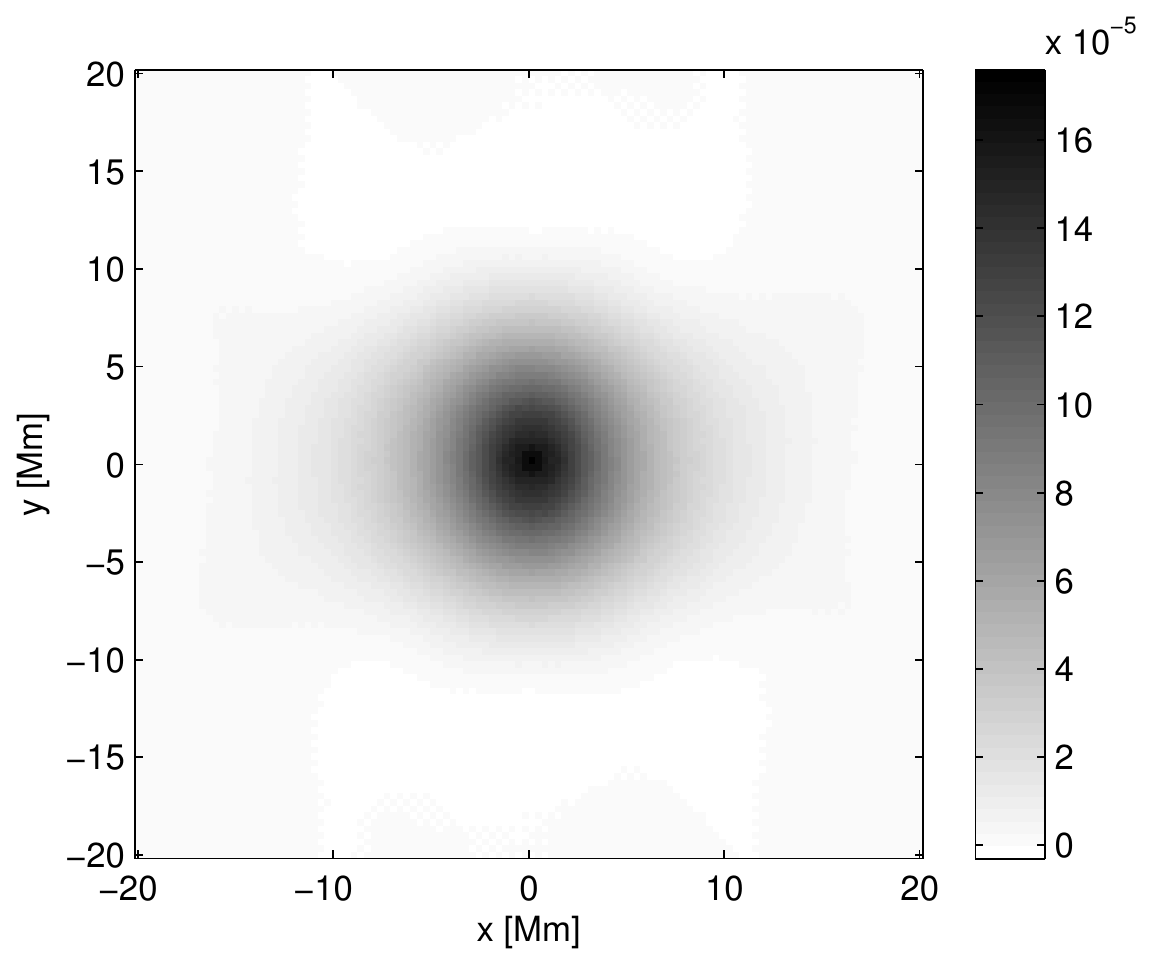}
               \hspace*{0.0\textwidth}
               \includegraphics[width=0.215\textwidth,clip=]{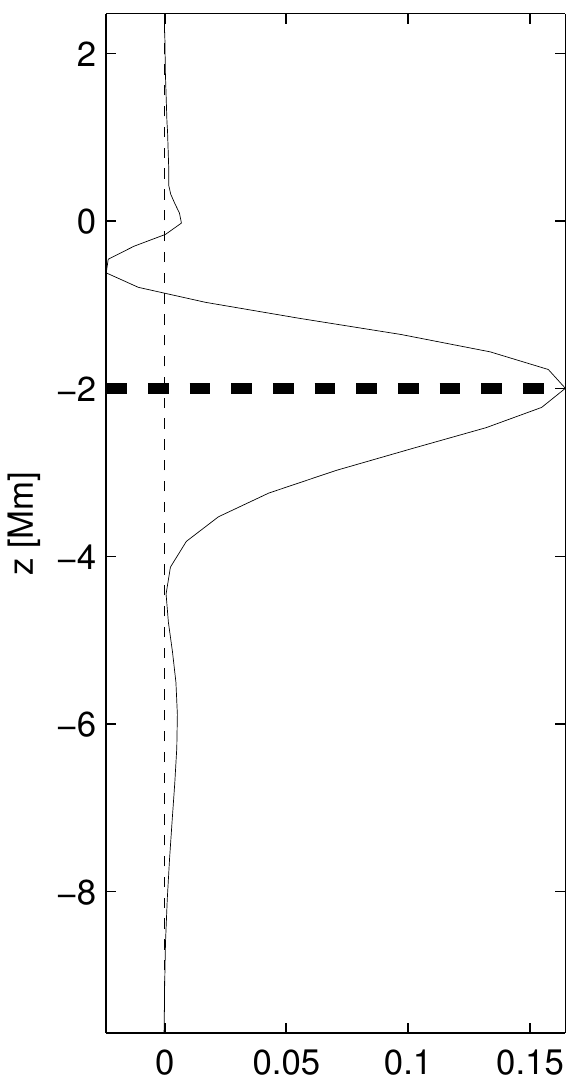}
              }
\caption{Averaging kernel for the $v_x$ inversion for $z=-2.0$ Mm.  The left plot shows the two-dimensional slice through the kernel at $z=-2.0$ Mm and the right plot shows the horizontally-integrated vertical distribution of the kernel.
}
\label{fig.kern_x}
\end{figure}

%
%
 Figure~\ref{fig.kern_x} shows the averaging kernel that relates the inferred $v_x$ to the true $x$-component of the velocity.  The RLS inversion procedure is able to produce reasonably localized $x$-component averaging kernels with minimal leakage outside the target region.  

\begin{figure}
\centerline{\includegraphics[width=5in,clip=]{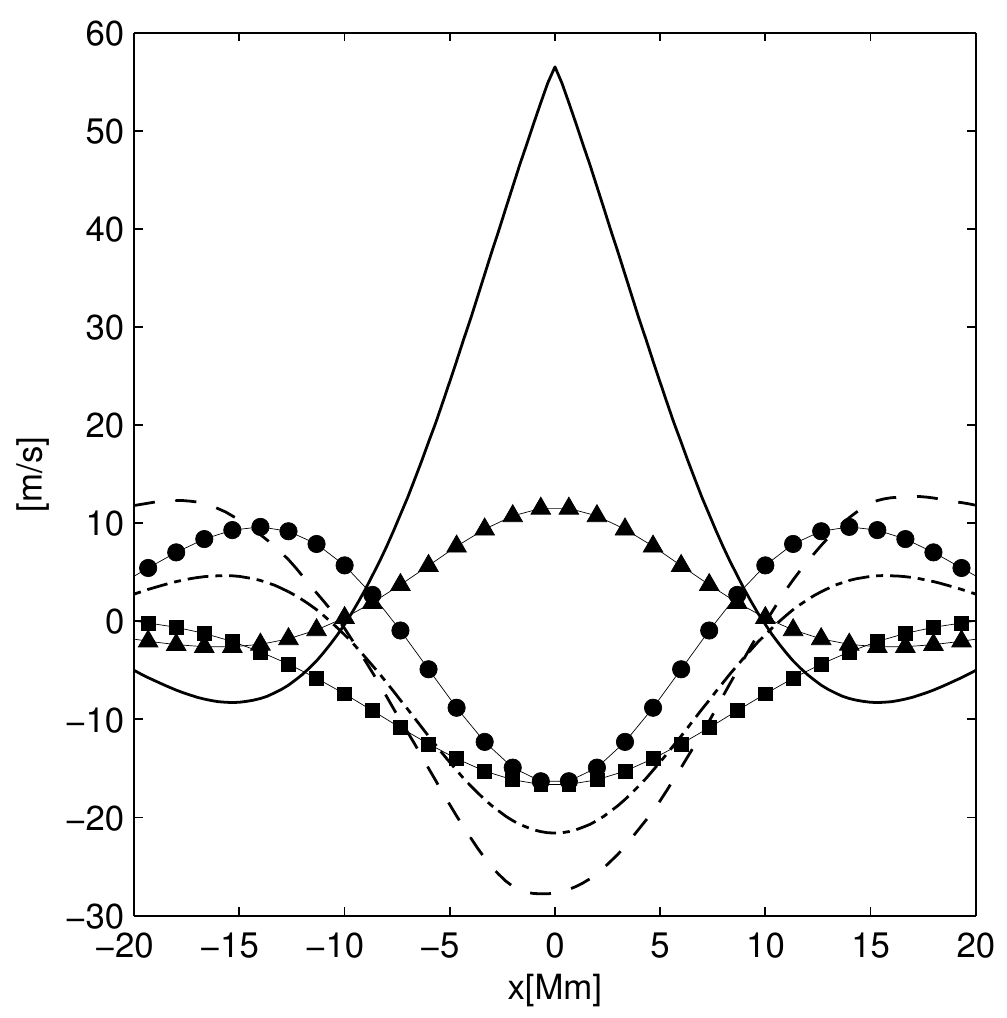}}
\caption{Convolutions between the $z$-directional kernel and components of the model flow [$v_{zx}$, squares; $v_{zy}$, circles; $v_{zz}$, triangles], the summation of the convolved components [$v_{\rm{tot}}= v_{zx}+v_{zy}+v_{zz}$, dot--dashed line], the inversion result [$v_z$ inv, dashed line], and the model supergranule flow [$v_z$ sim, solid line].  The $v_{zx}$- and $v_{zy}$-components have a large effect on the calculation of the total signal [$v_{\rm{tot}}$], which demonstrates that much of the error in the $v_z$-inversion can be attributed to cross-talk effects.}
\label{fig.conv_comp_z}
\end{figure}

%
%
\citet{Zhao2007} discuss the influence of cross talk on the inversions for the vertical velocity.  Specifically, they describe a region just below the photosphere where the $v_z$ inversion is of the incorrect sign.  We find a similar phenomenon in our results, illustrated by the band of negative vertical velocity near $z=0$ Mm (Figure~\ref{Vz_all}).  Here we will use the notation $v_{ij}$ to denote the contribution of the $j$~component of the (known) model velocity to the inversion result for the $i$~component of the velocity; computing each of the $v_{ij}$ involves the horizontal convolution of the appropriate averaging kernel with a particular component of the known model velocity \citep[see, \textit{e.g.}][for examples]{Jackiewicz2008}.   Figure~\ref{fig.conv_comp_z} shows the contributions of each of the components of the model flow to the inferred vertical flow (\textit{i.e.}~$v_{zx}$, $v_{zy}$, $v_{zz}$).  The contribution of the $x$- and $y$-components of the true flow to the inferred vertical flow are as large as the contribution from the true vertical flow.  The sum of these velocities ($v_{\rm{tot}} = v_{zx}+v_{zy}+v_{zz}$) is qualitatively similar to the vertical flow inferred from the inversion.  If the kernels were exactly correct and there was no noise then $v_{\rm tot}$ would be identical to the inferred vertical flow. This computation highlights the importance of controlling the cross-talk in inversions for velocity \citep[this has been achieved using OLA inversions by][]{Jackiewicz2008,Svanda2011}.

\section{Discussion}\label{sec.discussion}

We computed RLS inversions from travel-time measurements of simulated data created by numerically propagating waves through a simple supergranule-like flow field.  The benefit of this approach is the opportunity for comparison with a known flow field for direct evaluation of the performance of the measurement and inversion techniques.  Furthermore, the use of simulated data allows for isolation of complicating effects.  For instance, here travel-time differences are due solely to the supergranule flow; travel-time differences calculated from observational solar data contain the effects of many more physical variables.

We found that the inversion of the $v_x$- and $v_y$-velocity shared general features of the true flow reasonably well down to a depth of about 4 Mm below the surface.  The inversion of the vertical velocity performed poorly throughout the domain, particularly near the surface where the computations produced a result that was of incorrect sign.      

Underestimation of large-scale signal strength may be attributed to choice of regularization, which averages out noise but also causes some loss of flow amplitude.  We tested the inversions over a broad range of regularization parameters, and the results presented here reflect the best scenarios.  

We found the inversions, and hence the averaging kernels, to be relatively irresponsive to an increase in the number of input travel-time maps or to exclusion of noisy travel-time maps.  This is a credit to the capability of the RLS technique in minimizing the effects of noise; however, the significant error in the inversion results suggests that the forward modeling is inadequate.

The sensitivity kernels relate variations in travel-time shifts to variations in the solar model; capturing this sensitivity accurately is dependent on making appropriate assumptions regarding the physical state (in our case, the simulated state) and the measurement procedure \citep[\textit{e.g.}][]{Gizon2002, Birch2004}.  As a test of the kernels, we convolved the resulting averaging kernels with the components of the true supergranule flow field, and compared these results with the respective inversion results.  We found that the convolutions contained qualitatively the same structure (and hence comparable errors) as the inversion results.   

There is very little sensitivity to flows above the photosphere because only waves of frequency comparable to the acoustic cutoff frequency (about 5.5~mHz) sample this region.  The calculation of the kernels uses a zero Lagrangian pressure perturbation upper boundary condition, an assumption which is likely not accurate for high-frequency waves.  Furthermore, the simulation uses absorbing sponges at the top and bottom boundaries, which may affect the wave dynamics significantly differently from what is accounted for in the kernel calculations.  Perhaps for these reasons, the inversions are unable to reproduce the flow structure near the top of the computational domain.  Other constraints on the forward modeling and approximations used in the design of the sensitivity kernels may need to be considered in order to improve the performance of flow inversions.  

In conclusion, our findings indicate that errors in the RLS inversions result from the combined effects of cross talk between signals, choice of regularization parameters, design of sensitivity kernels, and inconsistencies among boundary conditions, all of which are open areas of research requiring further study. We note that improvement in performance, particularly in regard to cross-talk effects, has been seen in the optimally localized average (OLA) approach to reduce the size of the side lobes on the averaging kernels and limit spatial leakage \citep{Jackiewicz2008,Svanda2011,Jackiewicz2012}.

%
\begin{acks}
DED and ACB acknowledge support from NASA contracts NNH09CF68C and NNH07CD25C.  DCB acknowledges support from NASA contract NNH09CE41C.  The data from the simulations used in this study is located on a publicly accessible website (\url{http://www.cora.nwra.com/~dombroski/supergranulation/}) to facilitate further investigation. The authors thank the referee for helpful suggestions.
\end{acks}

\bibliographystyle{spr-mp-sola}

\end{article} 
\end{document}